\documentclass{PoS}
\usepackage[utf8]{inputenc}
\usepackage{amsmath}
\usepackage{amsfonts}
\usepackage{amssymb}
\usepackage{graphicx} 

\newcommand{\B}{\text{B}}
\newcommand{\done}{\text{d}}

\newcommand{\V}{\text{V}}
\newcommand{\R}{\text{R}}
\newcommand{\Sub}{\text{S}}
\title{The Electroweak Sudakov approximation in SHERPA}

\ShortTitle{The EW Sudakov approximation in SHERPA}

\author{\speaker{Jennifer M Thompson}\\%
        II. Physikalisches Institut\\
        University of Goettingen\\
        Friedrich-Hund-Platz 1\\
        37077 Göttingen\\
        E-mail: \email{jennifer.thompson@phys.uni-goettingen.de}}

\abstract{As experimental particle physics becomes more and more precise, it is becoming increasingly important for Monte Carlo simulations to improve the precision of their predictions. In terms of the hard matrix element, this means calculating to a higher order in perturbation theory. To be consistent this requires both NNLO QCD corrections and NLO EW corrections to be included. There are also interference effects between these processes that are not simple to handle consistently. For a broad description of the behaviour of NLO EW corrections at high energies, the Sudakov logarithmic approach provides a good approximation, and is much less computationally expensive than the full calculation. The implementation of EW Sudakov logarithms within the SHERPA program are outlined here along with some initial results. As well as this, an overview of the status of full NLO EW computations with SHERPA is presented.}

\FullConference{XXIV International Workshop on Deep-Inelastic Scattering and Related Subjects\\
                 11-15 April, 2016\\
                 DESY Hamburg, Germany}

\begin{document}

\section{Introduction}
Monte Carlo event generators provide an important bridge between collider experiments and the underlying physics of the SM. They enable low-multiplicity fixed-order calculations to be compared against the busy hadronic environment that characterises modern-day experimental particle physics. This is achieved by dressing the fixed-order calculation with a parton shower and introducing underlying events and hadronisation effects. The current state-of-the art for these predictions involves matching NLO QCD matrix elements to a parton shower. Moving beyond this perturbative accuracy requires either performing an NNLO QCD or an NLO EW calculation, both of which require new technology and often contribute a similar order of magnitude to the correction. There has been a lot of work on both calculating fixed order NNLO QCD corrections such as~\cite{Ridder:2013mf,Boughezal:2013uia,Czakon:2013goa} and on the matching of NNLO QCD cross sections with a parton shower~\cite{Hamilton:2012np,Hoche:2014dla,Hoeche:2014aia}. However, the work presented in the following considers the extension of Monte Carlo event generators, specifically the SHERPA~\cite{Gleisberg:2003xi,Gleisberg:2008ta} event generator, to include NLO EW corrections. There is a particular focus on the EW Sudakov approximation, the origin of which is outlined and the implementation within SHERPA briefly explained. Some initial results are presented for a 14 TeV LHC. 

\section{NLO QCD}

To begin a discussion on NLO EW, it is instructive to consider the comparative case of NLO QCD. Calculating the NLO QCD corrections to a process analytically is given by,

 \begin{equation}\label{EQN:NLOana}
  \sigma_{\text{NLO}}^{\text{QCD}} = \int(\B + \V)\done\Phi_\B + \int\R\done\Phi_\R\,,
 \end{equation}

which cannot be easily transferred to a numerical calculation due to the divergence of the virtual ($\V$) and real ($\R$) integrals on the RHS. The born contribution, $\B$, obviously does not contain any such divergences. Each term in eqn.\,(\ref{EQN:NLOana}) is integrated over its appropriate phase-space to give the full NLO cross section, $\sigma_{\text{NLO}}^{\text{QCD}}$. The solution to this problem which is almost universally adopted is to use a subtraction scheme~\cite{Catani:1996vz,Bevilacqua:2013iha,Frederix:2009yq,Campbell:1998nn,Kosower:2003bh}. This subtracts a quantity, $\Sub$ which exactly matches the divergent behaviour of the integral over the real emission. This must be analytically integrable over the 1 parton sub-space and not introduce any new divergences~\cite{Catani:1996vz}. Once these conditions are met, the subtraction term can be added to the virtual integral. Once it is appropriately integrated it will then also match exactly the divergent structure of the virtual contribution. With this introduction, eqn~(\ref{EQN:NLOana}) becomes a sum of finite integrals, 

\begin{equation}
\sigma_\text{NLO}^{\text{QCD}} = \int(\B+\V+\int\Sub\done\Phi_1)\done\Phi_\B+\int(\R-\Sub)\done\Phi_\R\,.
\end{equation}

\section{NLO EW}

\subsection{Performing the calculation}

There are several differences between an NLO QCD calculation and an NLO EW calculation. One obvious difference is the appearance of masses of the EW gauge bosons, which regulate the divergences from soft and collinear radiation. The real emission of $W^\pm$ and $Z$ bosons decay into other particles and are, theoretically, distinguishable processes from the underlying Born term. This reduces the problem of including real radiation to that of QED, which can be treated in a similar way to the NLO QCD calculation above. Because the real radiation can be, at least to a large extent, classified as a distinct process, the large corrections in the high energy regime, where the mass of the weak bosons becomes negligible, are physically meaningful. This high energy regime is the limit where the Sudakov logarithmic approximation becomes valid.

Another difference introduced in NLO EW calculations is the dependence of the EW bosons on the helicity of the particles involved, particularly clear in $W^\pm$ boson emission. Unlike in QCD, the couplings of the weak bosons are strongly dependent of the helicity of the particle. Furthermore, the exchange of weak particles can change the underlying Born term, for example mixing electrons with neutrinos, and therefore create interferences between previously distinct processes. These differences affect both the NLO EW calculation and the implementation of the EW Sudakov logarithmic approximation.

\subsection{EW Sudakov Approximation}

\begin{figure}
   \begin{center}
   \includegraphics[width=0.3\textwidth]{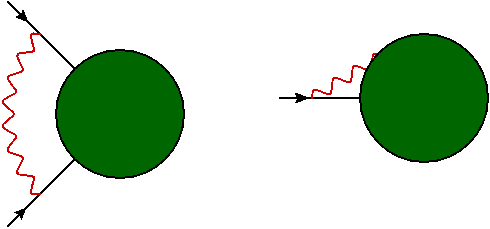}
   \caption{\label{FIG:mass} Diagrams which lead to mass singularities and contribute to the logarithmic approximation.}
   \end{center}
 \end{figure}

A full NLO EW calculation is very computationally intensive, whereas the EW Sudakov approximation does not introduce much overhead and is therefore comparatively cheap. It is also easier to include on top of NLO QCD calculations. This approximation considers only the logarithmic contribution to the correction, and is dominant in the high-energy regime.
The EW Sudakov approximation in SHERPA follows an excellent and clear paper by Denner and Pozzorini~\cite{Denner:2000jv}, which provides a break down of how to implement Sudakov logarithms in a process-independent way. These logarithms, $L$, typically take the form

\begin{equation}\label{EQN:sud1}
L\sim\log\left(\frac{s}{M_V^2}\right)\,,
\end{equation}

which captures the high energy behaviour of the NLO EW calculations. In eqn\,(\ref{EQN:sud1}), $s$ is the centre of mass energy, and $M_V$ is the mass of the relevant weak boson. There is always a non-logarithmic piece which is neglected, and this creates a limit to the accuracy EW Sudakov logarithms can reach, typically $\sim 1\%$. These logarithms are produced by the exchange of soft-collinear weak bosons, which become divergent in the limit of vanishing mass, and are referred to as mass singular diagrams, as discussed in ref.~\cite{Kinoshita:62}. These diagrams involve either the exchange of an EW boson between 2 external legs (double logarithms) or the emission of a soft or collinear boson from an external leg (single logarithms). These contributing diagrams are shown in fig.\,\ref{FIG:mass}. The right-hand diagram in fig.\,\ref{FIG:mass} includes the wave-function renormalisation terms where the boson is reabsorbed by the emitting line. There are also single logarithms from parameter renormalisation, which are not depicted.

It is clear that the double logarithmic terms are the dominant contribution, with the single logarithms providing a subleading contribution. However, while the double logarithms are typically a negative correction to the Born cross section, the subleading contribution is a positive correction. Considering only the leading contribution therefore leads to an overestimation of the size of the Sudakov correction.

In the one-loop approximation, the EW Sudakov logarithms introduce corrections, summed over external legs $i$ and $j$ and EW bosons with typical mass $M$,

\begin{equation}\label{EQN:sudfin}
   L=\frac{\alpha}{4\pi}\left[A\log^{2}\left(\frac{(p_i+p_j)^2}{M^2}\right)
   +B\log\left(\frac{(p_i+p_j)^2}{M^2}\right)\right]\,,
\end{equation}

where $p_i$ denotes the momentum of leg $i$.

\subsection{EW Sudakov results}

Within the SHERPA framework, the EW Sudakov approximation is included as a K-factor that is applied at each phase-space point. As is implied by eqn~(\ref{EQN:sudfin}), it depends only on the final state legs, and iterates over all possible exchanges of EW bosons. All bosons are assumed to have a mass equal to the $W^\pm$ boson mass. This introduces only a small logarithmic correction for the $Z$ boson, which can be neglected to the order considered here. The mass difference between the photon and the $W^\pm$ boson introduces large logarithms, but these largely cancel against real photon radiation. Therefore, the assumption that the EW bosons all have equal mass does not have a significant impact on the correction. The EW Sudakov approximation only affects the hard process and can be easily employed with the parton shower or applied to an NLO QCD computation. However, the implementation does rely on the COMIX~\cite{Gleisberg:2008fv} matrix element generator. 

\begin{figure}
   \begin{center}
   \includegraphics[width=0.4\textwidth]{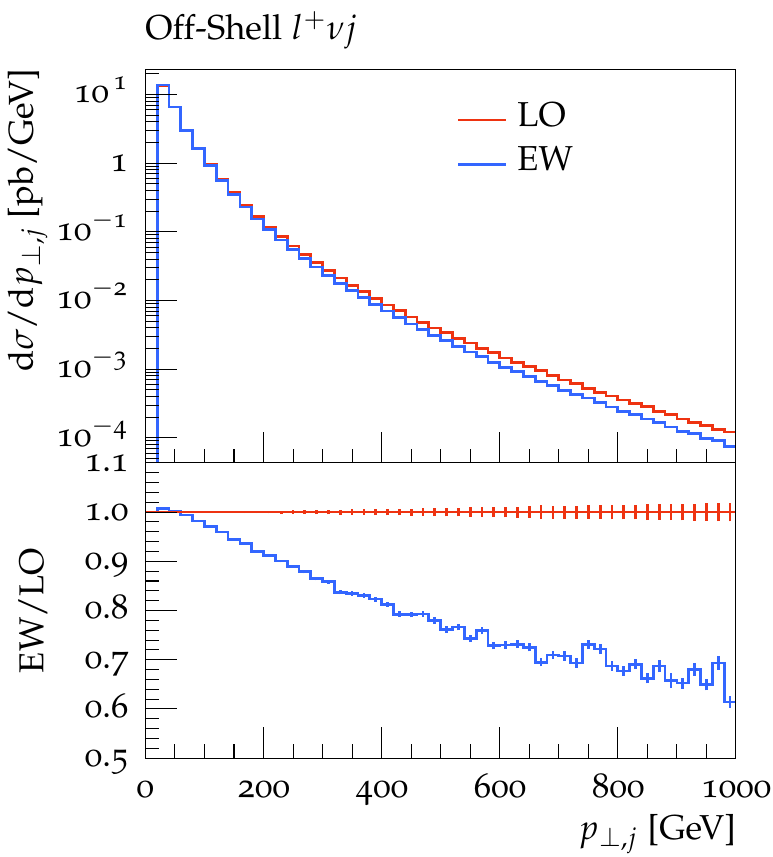}
   \includegraphics[width=0.4\textwidth]{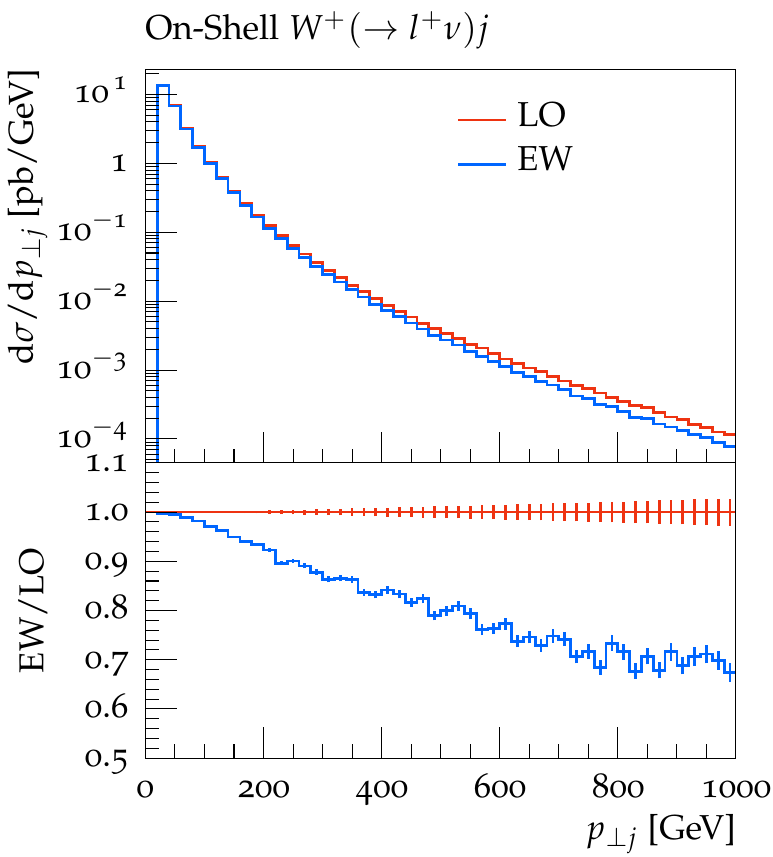}
   \caption{\label{FIG:onoff} $p_T$ of the leading jet in off-shell (left-hand side) and on-shell (right-hand side) production of a $W^\pm$ boson with a jet.}
   \end{center}
 \end{figure}

The results shown here are the first results with the implementation of NLO EW Sudakov logarithms within SHERPA. The calculations are performed for a 14 TeV LHC collider.
The left-hand side of fig.\,\ref{FIG:onoff} shows the effect of including NLO EW Sudakov corrections at 14 TeV in off-shell $W^\pm$+jets production. It is clear that as the $p_T$ of the leading jet increases, the relative correction from the EW Sudakov logarithms becomes larger and more negative. This reaches almost $40\%$ at 1 TeV. The behaviour is similar on the right-hand side of fig.\,\ref{FIG:onoff}, throughout the $p_T$ spectrum, which shows the same distribution but for on-shell production of the $W^\pm$ boson.

\subsection{Full NLO EW}

Although the NLO EW Sudakov approximation is comparatively quick and easy to include on top of NLO QCD computations, for some studies, either to a much higher precision or outside of the high energy regime, a full NLO EW computation must be employed. This faces the same challenges and subtleties that the NLO EW Sudakov approach dealt with, alongside ambiguities in process definition. It must be decided, for example, what counts as a photon emitted in the NLO EW calculation and what is simply radiation from a jet. Also, the difference between NLO QCD corrections to EW processes and NLO EW corrections to QCD processes must be defined in order to avoid double counting.

Within the SHERPA event generator, there is currently a working interface to the OpenLoops~\cite{Cascioli:2011va} loop provider for the NLO EW virtual amplitude and QED subtraction handled within SHERPA. There are already publications on NLO EW corrections, including multijet merging, to $V$+jets~\cite{Kallweit:2014xda,Kallweit:2015dum}, both with an on-shell $W^\pm$ boson and including off-shell effects, however the code is not yet public. There is also ongoing effort to implement an NLO EW interface between SHERPA and Recola~\cite{Actis:2016mpe}, as another one-loop provider for both NLO QCD and NLO EW virtual amplitudes. 

\section{Conclusions}

Improving the perturbative accuracy of the hard interaction in Monte Carlo event simulation now involves the calculation of NLO EW contributions, which are often of a comparable size to NNLO QCD. This includes several new challenges, and the full computation is quite time-intensive. EW Sudakov logarithms provide a simple way for the dominant behaviour of the NLO EW calculation to be taken into account without the computational overhead. It is also trivial to include on top of QCD corrections, unlike the full calculation where the interference terms must be carefully considered to avoid double counting. This is an important complementary approach to be implemented alongside the full calculation. There is also a lot of promising progress in the automated evaluation of the full NLO EW correction.

\begin{acknowledgments}
This work has been supported by the European Commission through the 
networks MCnetITN (PITN--GA--2012--315877).
\end{acknowledgments}

\end{document}